\documentclass[apjl,twocolumn]{aastex631}

\newcommand{\rev}\textbf{}
\newcommand{\revm}\mathbf{}

\newcommand{\kms}{\,km\,s$^{-1}$}

\newcommand{\degree}{$^{\circ}$}
\newcommand{\Rstar}{$\mathrm{R_{\star}}$}

\newcommand{\TBJD}{$\mathrm{BJD_{TDB}}$}
\newcommand{\REarth}{$\mathrm{R_{\oplus}}$}
\newcommand{\MEarth}{$\mathrm{M_{\oplus}}$}
\newcommand{\RJup}{$\mathrm{R_{J}}$}
\newcommand{\MJup}{$\mathrm{M_{J}}$}

\usepackage{comment}
\usepackage{soul}
\shorttitle{The obliquity of HIP67522 b}
\shortauthors{Heitzmann et al.}

\graphicspath{{./}}

\begin{document}

\title{The obliquity of HIP 67522 b: a 17\,Myr old transiting hot Jupiter-sized planet}

\correspondingauthor{Alexis Heitzmann}
\email{alexis.heitzmann@usq.edu.au}

\author[0000-0002-8091-7526]{Alexis Heitzmann}
\affiliation{Centre for Astrophysics, University of Southern Queensland, West Street, Toowoomba, QLD 4350 Australia}

\author[0000-0002-4891-3517]{George Zhou} 
\affiliation{Centre for Astrophysics, University of Southern Queensland, West Street, Toowoomba, QLD 4350 Australia}

\author[0000-0002-8964-8377]{Samuel N.~Quinn} 
\affiliation{Center for Astrophysics \textbar{} Harvard \& Smithsonian, 60 Garden St., Cambridge, MA 02138, USA.}

\author[0000-0001-5522-8887]{Stephen C.~Marsden} 
\affiliation{Centre for Astrophysics, University of Southern Queensland, West Street, Toowoomba, QLD 4350 Australia}

\author[0000-0001-7294-5386]{Duncan Wright}
\affiliation{Centre for Astrophysics, University of Southern Queensland, West Street, Toowoomba, QLD 4350 Australia}

\author[0000-0001-7624-9222]{Pascal Petit}
\affiliation{Institut de Recherche en Astrophysique et Plan\'{e}tologie, Universit\'{e} de Toulouse, CNRS, CNES, 14 avenue Edouard Belin, 31400 Toulouse, France}

\author[0000-0001-7246-5438]{Andrew M.~Vanderburg}
\affiliation{Department of Physics and Kavli Institute for Astrophysics and Space Research, Massachusetts Institute of Technology, 77 Massachusetts Avenue, Cambridge, MA 02139, USA}

\author[0000-0002-0514-5538]{Luke G.~Bouma}
\affiliation{Department of Astrophysical Sciences, Princeton University, NJ 08544, USA.}

\author[0000-0003-3654-1602]{Andrew W.~Mann}
\affiliation{Department of Physics and Astronomy, The University of North Carolina at Chapel Hill, Chapel Hill, NC 27599, USA}

\author[0000-0001-9982-1332]{Aaron C.~Rizzuto}
\affiliation{Department of Astronomy, The University of Texas at Austin, Austin, TX 78712, USA}

\begin{abstract}

HIP 67522 b is a 17 Myr old, close-in ($P_{orb} = 6.96$ d), Jupiter-sized ($R = 10$ \REarth) transiting planet orbiting a Sun like star in the Sco-Cen OB association. We present our measurement of the system's projected orbital obliquity via two spectroscopic transit observations using the CHIRON spectroscopic facility. We present a global model that accounts for large surface brightness features typical of such young stars during spectroscopic transit observations. With a value of $|\lambda| = 5.8^{+2.8}_{-5.7}$\degree, it is unlikely that this well-aligned system is the result of a high eccentricity driven migration history. By being the youngest planet with a known obliquity, HIP 67522 b holds a special place in contributing to our understanding of giant planet formation and evolution. Our analysis shows the feasibility of such measurements for young and very active stars.

\end{abstract}


\section{Introduction} \label{sec:intro}

One of the oldest puzzles in the field of exoplanets is the origin of short-orbit gas giants. With no equivalent in the Solar system, it is far from obvious how these giant planets, orbiting their host star incredibly close (with orbital periods $< 10$ days) come to exist.

Among the observational properties easily measurable for these exoplanets when they transit, their sky-projected stellar obliquity ($\lambda$) angles may help differentiate between the multiple pathways explaining their migration \citep{2018ARA&A..56..175D}. Unfortunately, on its own, the obliquity angle can not unambiguously identify a specific formation pathway. Star-planet tidal interactions, resulting in angular momentum exchanges between a host star and its planet, have the ability to circularize and shrink planetary orbits but also alter the star's rotation axis alignment. This can erase primordial orbital characteristics (i.e. misalignments and/or eccentricity), preventing the identification of a specific migration channel. Because the timescales of such interactions span many orders of magnitudes (from $10^5$ to $10^9$ years), the $\sim 150$\footnote{TEPCat July 2021 \citealt{2011MNRAS.417.2166S}} sky-projected orbital obliquity measurements obtained to date remain difficult to interpret \citep{2018haex.bookE...2T,2021ApJ...916L...1A}. 

Recently, it became possible to compare the obliquity distribution for very young stars ($< 100$ Myr) against that of more mature stars. These young planetary systems have yet been influenced by star-planet tidal effects, and provide a glimpse into the primordial orbits of planets post-formation. The recent efforts to characterize young planets discovered by the \emph{K2} and \emph{Transiting Exoplanet Survey Satellite} (TESS, \citealt{2016SPIE.9904E..2BR}) missions, such as AU Mic b,  \citep{2020A&A...643A..25P,2020arXiv200613675A,2020A&A...641L...1M,2020ApJ...899L..13H}, V1298 Tau c,  \citep{2019ApJ...885L..12D,2021arXiv210701213F}, DS Tuc Ab, \citep{2020ApJ...892L..21Z,2020AJ....159..112M} and TOI 942 b, \citep{2021arXiv210614968W,2021AJ....161....2Z},
have the potential to deliver key insights on the formation and migration of close-in planets. Although these transiting planets orbiting rapidly rotating young stars are suitable candidates for obliquity measurements, their host stars' young age imply that strong intrinsic variability needs to be dealt with in order to recover the true spin-orbit angles.

In this letter we present a projected obliquity measurement for HIP 67522 b. This obliquity measurement is the first for a hot, Jupiter-sized planet younger than 100 Myr. With an age of only 17 Myr, HIP 67522 b has a radius of $10.178\pm0.440$ \REarth\, (this work), a mass $<5$ \MJup\,\citep{2020AJ....160...33R} and is orbiting a bright ($\mathrm{V}_{mag} = 9.876 \pm 0.026$) Sun-like star ($\mathrm{T_{eff}} = 5675\pm75$\,K). 

We describe both photometric and spectroscopic observations of HIP 67522 b's transit in Section~\ref{sec:obs} and present our combined model used to determine the projected obliquity of the system in Section~\ref{sec:analysis}. Finally, in Section~\ref{sec:discussion}, we place this measurement into context of other planetary systems around mature main-sequence stars. 

\section{Observations} \label{sec:obs}

\subsection{TESS: Photometry} \label{subsec:tess}

\begin{figure*}
    \centering
    \includegraphics[width=1\linewidth]{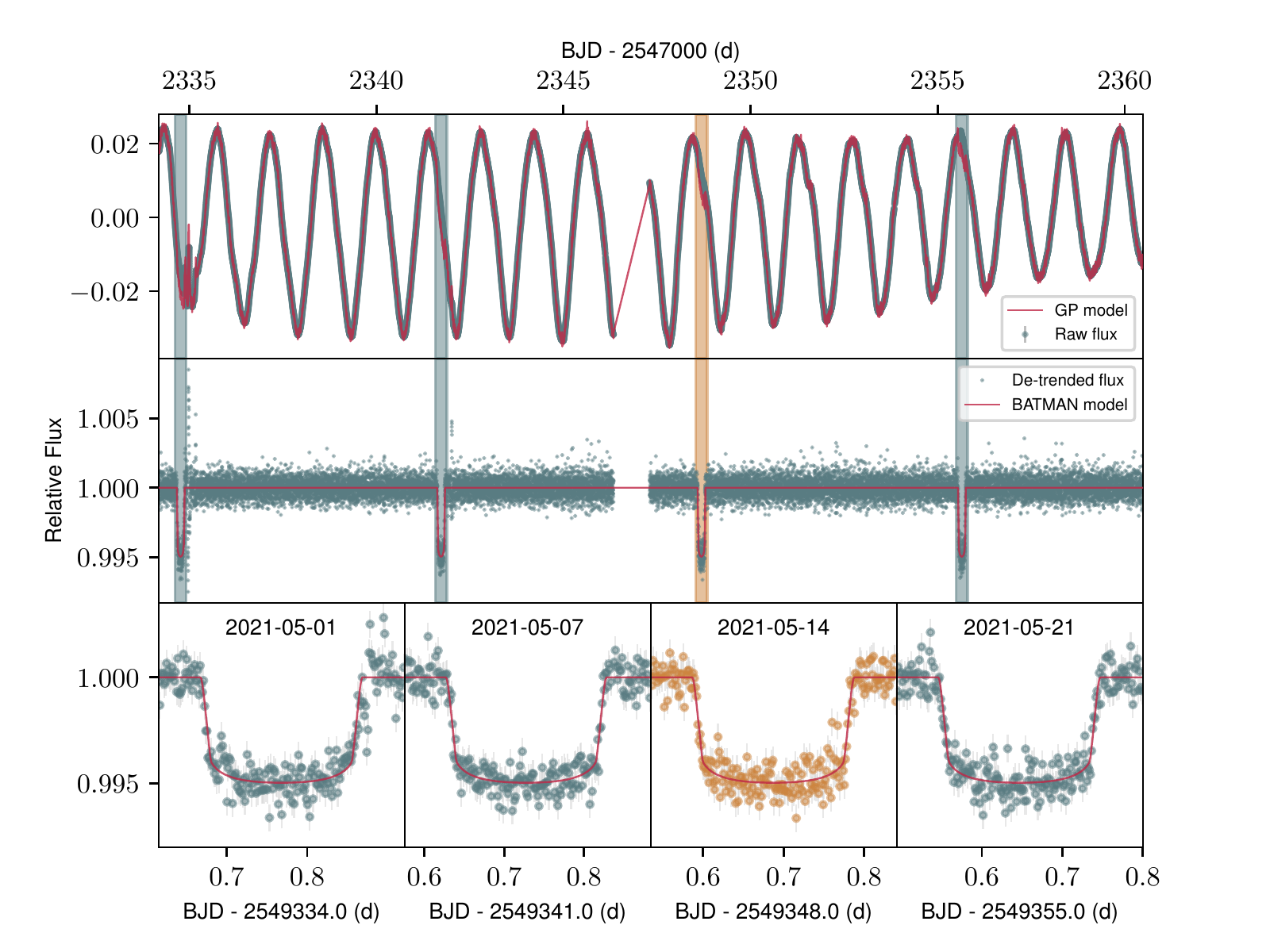}
    \caption{\textbf{Top:} TESS light curve for Sector 38. Data points are shown as gray points. The red line demonstrates the Gaussian Process model describing the stellar variability of the TESS light curves. Locations of individual transit events are highlighted by the shaded regions. \textbf{Middle:} The TESS Sector 38 light curve after the Gaussian Process model has been removed. \textbf{Bottom: } Close-ups of each transit event. The transit event on 2021-05-14 was simultaneously observed by CHIRON, and is marked in orange for clarity.}
    \label{fig:lightcurve}
\end{figure*}

HIP 67522 b was first identified to transit in Sector 11 of the TESS primary mission \citep{2020AJ....160...33R} over the period of 2019-04-22 to 2019-05-21. The target was subsequently observed by TESS again during Sector 38 of the extended mission, over the period of 2021-04-28 to 2021-05-26. We make use of the Simple Aperture Photometry \citep{twicken:PA2010SPIE,morris2020} made available for the target star extracted by the Science Processing Observation Center \citep[SPOC,][]{2016SPIE.9913E..3EJ} from the target pixel files, obtained at 2 minute cadence from both sectors of observations. The light curves were detrended against spacecraft motion via the quaternion detrending technique as per \citet{2019ApJ...881L..19V}, allowing the recovery of two missing transits in the Sector 11 observations as shown in the discovery paper. Except for these two transits, the light curve was also decorrelated against the PDC band 3 (fast timescale) cotrending basis vectors, and modeled the rotation signal with a basis spline with breakpoints every 0.1 days while excluding points with transits from the systematics correction. This allowed the recovery of a systematic corrected light curve from Sector 38. The light curve from Sector 38 is shown in Figure~\ref{fig:lightcurve}. 

\subsection{CHIRON/SMARTS: Spectroscopy} \label{subsec:chiron}

We obtained two spectroscopic transits of HIP 67522 b with the CHIRON facility \citep{2013PASP..125.1336T}. CHIRON is a high resolution echelle spectrograph on the 1.5\,m Small and Moderate Aperture Research Telescope System (SMARTS) telescope, located at Cerro Tololo Inter-American Observatory, Chile. CHIRON is fed via a fiber bundle, and has a spectral resolving power of $\lambda / \Delta \lambda \equiv R \approx 80,000$ with a wavelength coverage from 4100 to $8700$\,\AA{}. 

A total of 24 observations, with exposure times of 1200s each, were obtained on 2021-05-14, capturing the full transit and baselines on pre-ingress and post-egress from 2021-05-14 00:00 to 08:00 UTC. An additional partial transit was observed from 2021-06-17 23:15 to 2021-06-18 05:20 UTC, with a total of 18 observations at 1200s integration time each. The stellar spectra were extracted via the official CHIRON pipeline (Paredes et al 2021, submitted), with wavelength calibration provided by a set of Thorium-Argon cathode ray lamp exposures that bracket each transit sequence. To derive line broadening profiles from each spectrum, we perform a least-squares deconvolution of the observed spectrum against a set of synthetic non-rotating spectral templates \citep[][]{1997MNRAS.291..658D,2010MNRAS.407..507C}, making use of the ATLAS9 atmosphere models \citep{Castelli:2004} computed at the atmospheric parameters of HIP 67522 (Teff = 5725 K, log $g$ = 4.0 and $\left[\mathrm{Fe}/\mathrm{H}\right]$ = 0). To ensure realistic uncertainties on our measurement of lambda, we re-binned these line profiles to the velocity dispersion corresponding to the detector's pixel size. We make use of these for the transit spectroscopic modeling as described in Section~\ref{subsec:transitsspectro}. In addition, we also model each line profile with a kernel that incorporates the effects of rotational, macroturbulent, and instrumental line broadening. We make use of this model to determine the rotational broadening velocity $v\sin\,i_\star$ necessary for the line profile modeling, measuring a rotational broadening velocity of $v\sin\,i_\star=50\pm3\,$\kms, consistent with that of $54.2 \pm0.7$\,\kms reported in the discovery paper from an ensemble of spectroscopic observations. 

\section{Analysis} \label{sec:analysis}

The very young age of HIP 67522 goes hand in hand with substantial intrinsic stellar variability in both spectroscopic and photometric observations, with 2-3\% variations seen in both Sectors 11 and 38 TESS light curves. The variability seen for HIP 67522 can be mostly attributed to surface brightness features (spots and plages). We develop a model below that incorporates both photometric (Section~\ref{subsec:transitsphoto}) and spectroscopic transits (Section~\ref{subsec:transitsspectro}) as well as spot modeling to deal with the influence of stellar activity on our projected obliquity measurements.

\subsection{Transit photometry} \label{subsec:transitsphoto}

A total of eight transits, four in Sector 11 and four in Sector 38, were observed by TESS. The TESS photometry exhibits significant stellar rotational variability due to the youth of the host star. We apply a Gaussian Process model to account for this variability. We model the light curve using a stochastically drive Simple Harmonic Oscillation kernel as is implemented in the \textsc{celerite} package \citep{2017AJ....154..220F}, with free parameters describing its amplitude $S_0$ and damping coefficient $Q$. We fixed the frequency $\omega_0 = 1/P_{rot} = 1/1.39$ (see last paragraph of Section~\ref{subsec:transitsspectro} for justification). We also simultaneously model the planetary transit using the \textsc{batman} package \citep{2015PASP..127.1161K}, including free parameters describing the transit centroid timing $T_{\mathrm{c}}$, orbital period $P_{\mathrm{orb}}$, radius ratio $R_{\mathrm{p}}$/$R_{\mathrm{\star}}$, normalized semi-major axis $a$/$R_{\mathrm{\star}}$, and the line of sight inclination of the transit $i$. Following the low eccentricity found in \citet{2020AJ....160...33R}, the orbit is assumed to be circular for this model. 

The light curve and best fit model for the Sector-38 observations are shown in Figure~\ref{fig:lightcurve}. We subtract the Gaussian Process model from the TESS observations, and pass the resulting detrended light curve to the subsequent analysis described in Section~\ref{subsec:modeltotal}.

\subsection{Transit spectroscopy} \label{subsec:transitsspectro}

Any phenomenon introducing brightness variations on the stellar disk that are asymmetric will leave rotationally modulated imprints on the observed spectroscopic line profiles. As these variations move across the stellar disk, they block/enhance incoming flux at a wavelength (and corresponding radial velocity) depending on their longitude. This is because light emitted by different parts of the stellar disk experience a different Doppler shift due to the star's rotation. This directly translates into bumps and dips on the rotational line profiles. Doppler Tomography consists in monitoring the evolution of line profiles during a planetary transit to catch the distortions induced by the eclipsing body, called the 'Doppler shadow'. The way this Doppler shadow crosses the line profiles over time yields the angle at which the planet crosses the stellar disk, the projected spin-orbit alignment $\lambda$. Any other brightness variations (i.e. spot-induced features in a first approximation) will contribute to distort the line profiles in a similar fashion. We therefore modeled both contribution (spots and planet) to fit the observed line profiles and recover $\lambda$. \\
The star was modeled as a circular mask on a grid of pixels, each with a value between 0 and 1 representing fractional brightness $f_{\mathrm{b}}$. The mask is a combination of (i) a uniform disk ($f_{\mathrm{b}}$ = 1), (ii) a Limb Darkening quadratic law parametrized with a linear $\nu_1$ and a quadratic $\nu_2$ coefficient (given in Table~\ref{tab:priors}), (iii) the spot(s) and (iv) the transiting planet. Spots were modeled as spherical caps defined by their co-latitude (complementary angle of latitude, i.e. 0$^\circ$ at the north pole and 180$^\circ$ at the south pole) $\phi$, longitude $\theta$, angular radius $R_{\mathrm{spot}}$ and temperature $T_{\mathrm{spot}}$ conditioning their brightness, as per
\begin{equation}
    f_{\mathrm{b}} = \left(T_{\mathrm{spot}}/T_{\mathrm{eff}}\right)^4 \, .
\end{equation}
The projection of each spot on the stellar disk was then computed analytically and its contribution was added to the total mask. Finally, the transiting planet was assumed on a circular orbit and modeled as a disk ($f_b = 0$). Its shadow on the stellar disk was parametrized by $T_{\mathrm{c}}$, $P_{\mathrm{orb}}$, $R_{\mathrm{p}}$/$R_{\mathrm{\star}}$, $a$/$R_{\mathrm{\star}}$, $\omega$, $i$ and  $\lambda$ (see values in Table~\ref{tab:priors}). 

To obtain the full disk-integrated line profiles, we summed the pixels along the vertical axis of the stellar disk (i.e. for each radial velocity bins). Finally, we incorporate line broadening in the observations via a convolution of our model with a Gaussian kernel, with the width being the quadrature addition of the instrumental resolution and the macroturbulent velocity ($v_\mathrm{macro}$). 

\begin{figure*}    
    \gridline{\hspace{1cm}\fig{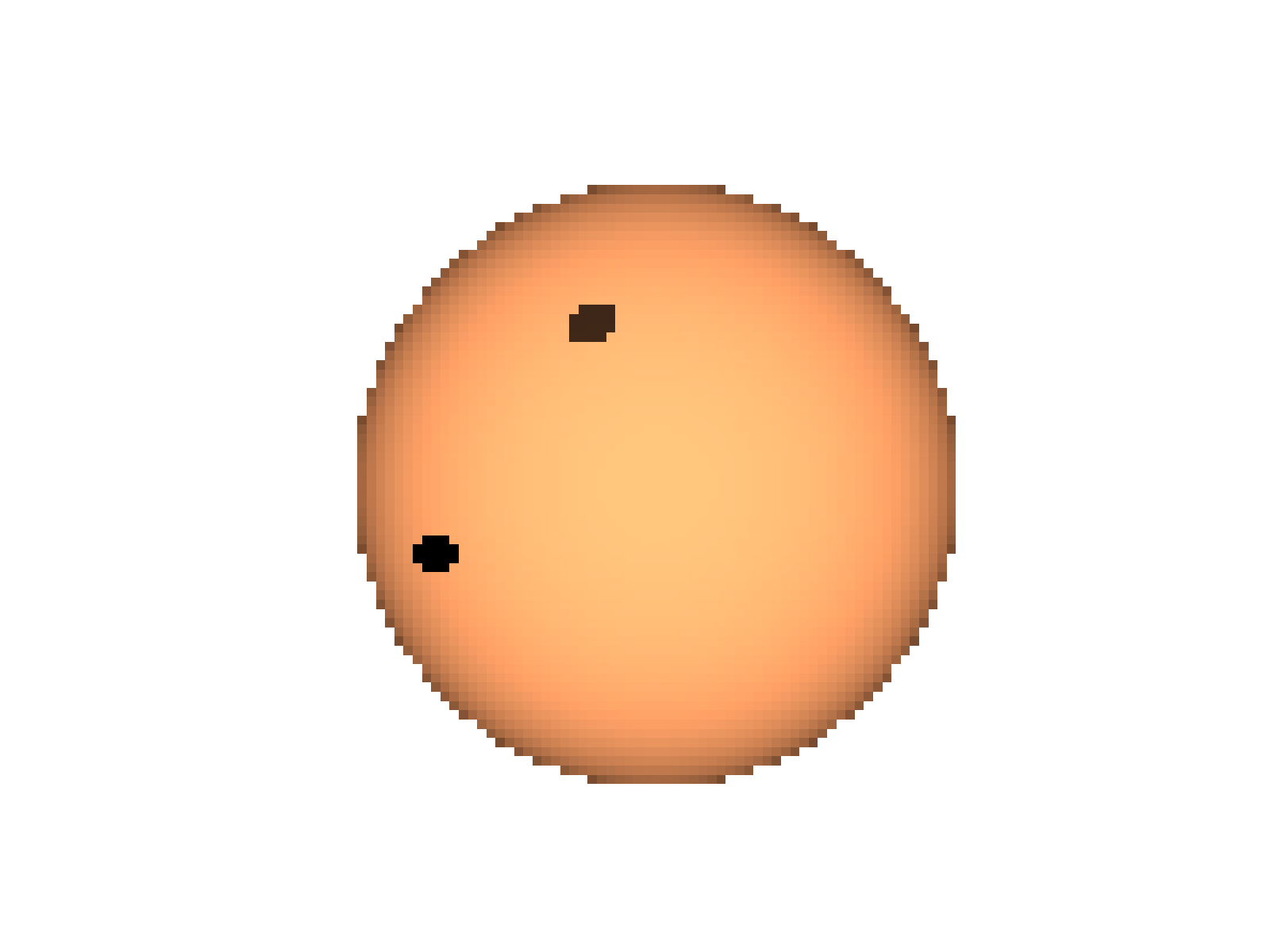}{0.25\textwidth}{}
              \hspace{-1.2cm}\fig{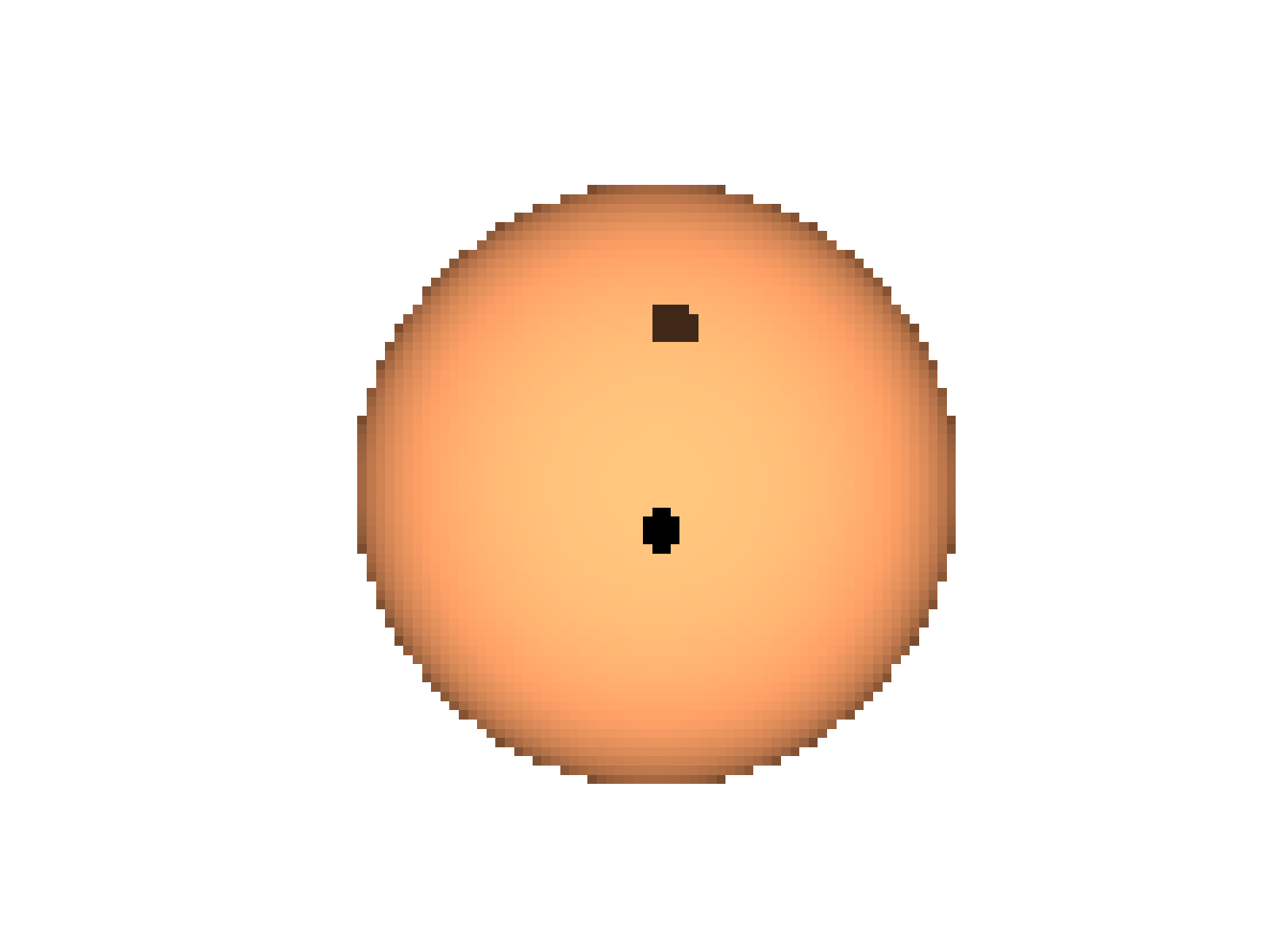}{0.25\textwidth}{}
              \hspace{-1.2cm}\fig{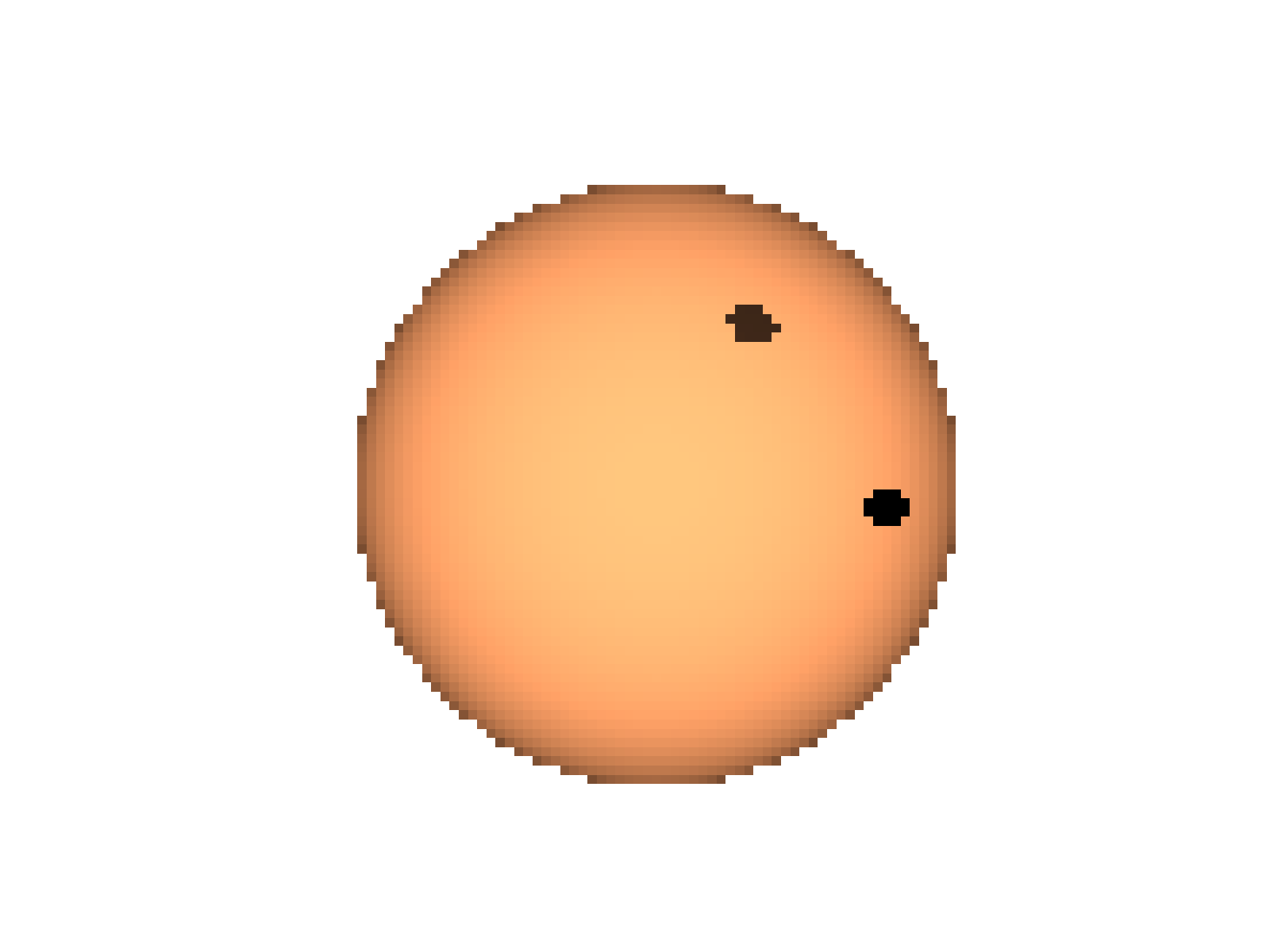}{0.25\textwidth}{}
              \hspace{1cm}}
    \gridline{\fig{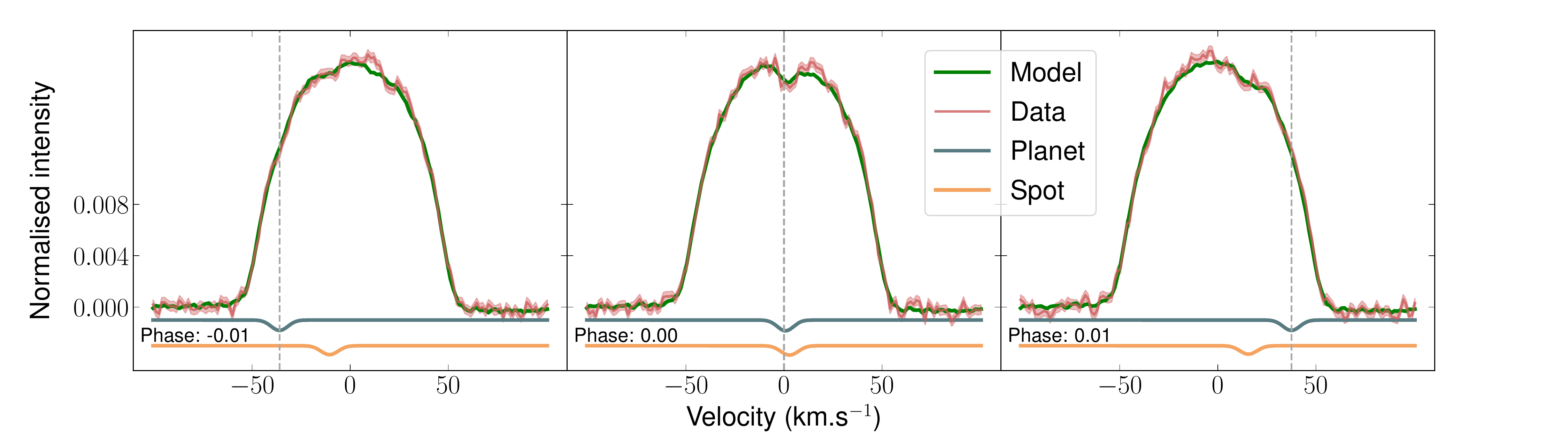}{\textwidth}{}}
    \gridline{\fig{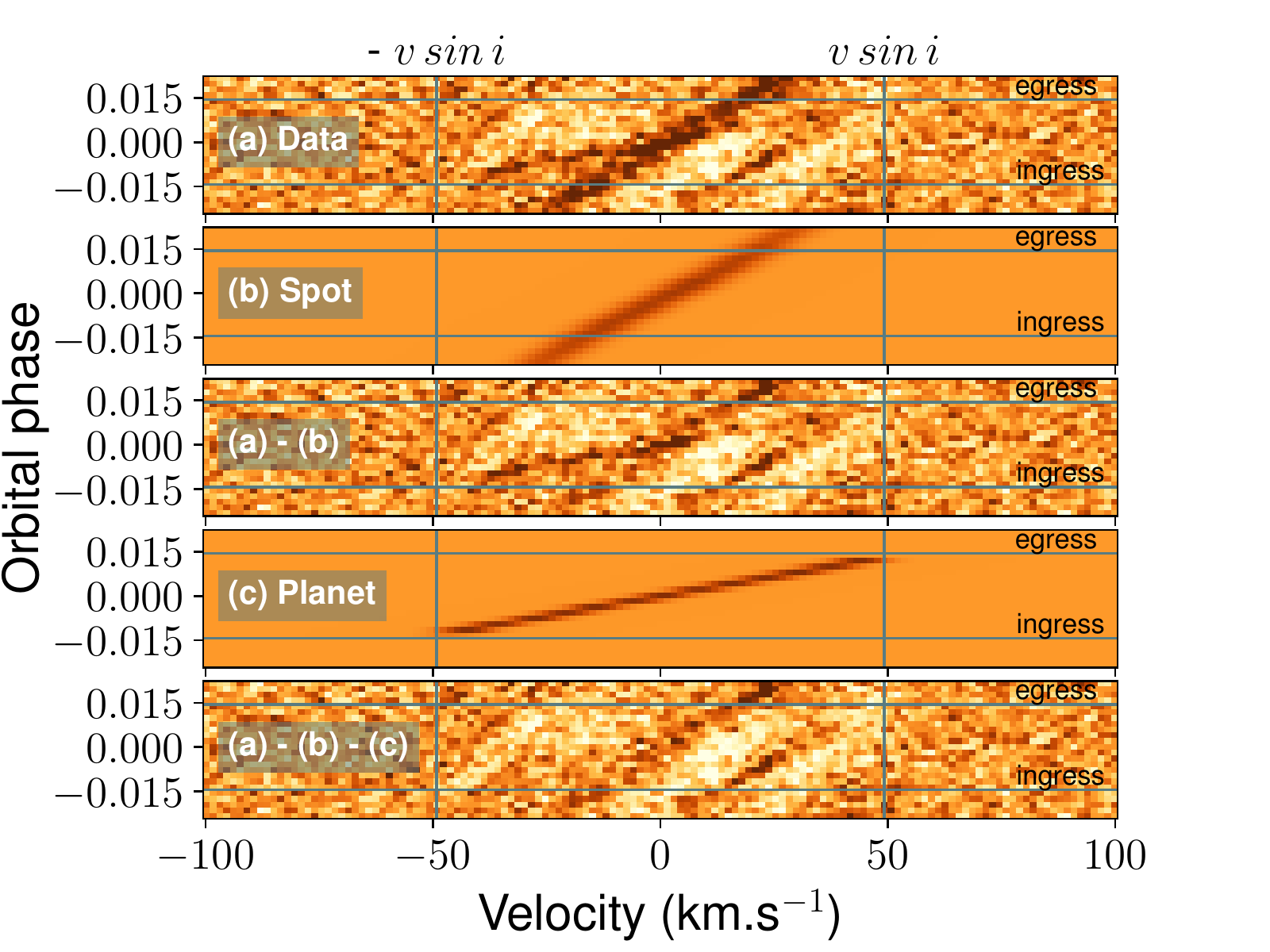}{0.48\textwidth}{2021-05-14 full transit.}
          \fig{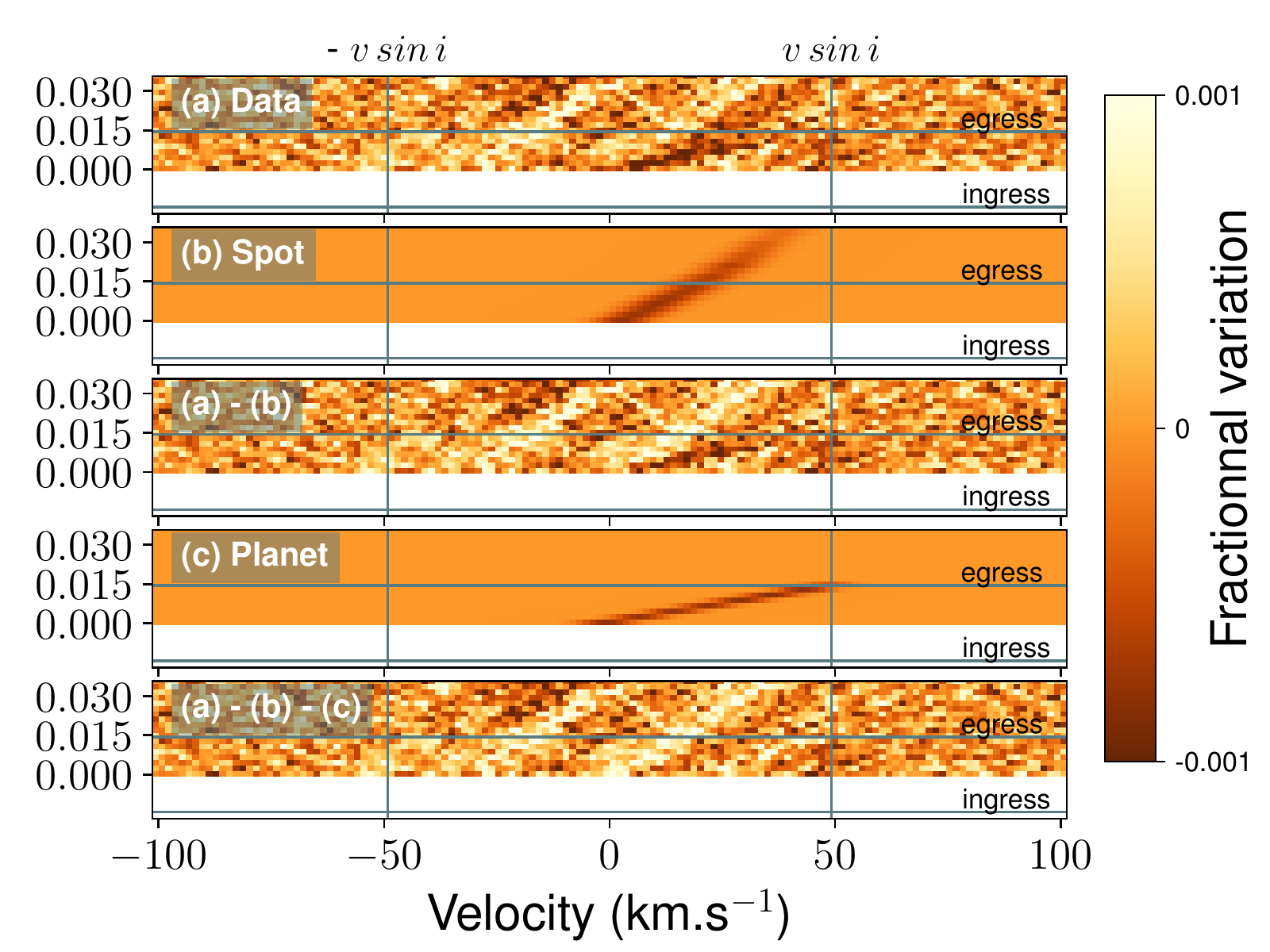}{0.48\textwidth}{2021-06-18 partial transit.}}
        \caption{Results from the spectroscopic transit modeling of HIP 67522 b. \textbf{Top}: Modeled stellar disk including a single spot and the transiting planet at phase -0.01 (left), 0 (center) and 0.01 (right). \textbf{Middle}: Line profiles fit matching the above images. Observed line profiles are shown in red with uncertainties in lighter red. Fit to the data (mean line profile + integrated modeled stellar disk) is shown in green, the planet contribution in blue and the spot contribution in orange. Gray vertical dotted lines express planet transit phases. \textbf{Bottom}: Result plots for the 2021-05-14 full transit (left) and the 2021-06-18 partial transit (right). Each subplot shows the variation of lines profiles in velocity space (horizontal axis) for different orbital phases of HIP 67522 b (vertical axis). (a): residuals between each observed line broadening and the mean observed line profile. (b): spot-only model. (a) - (b): line profile residuals after subtraction of the spot model, we see that the spot signature has been removed, leaving the clear planetary signature. (c): modeled planet. (a) - (b) - (c): residuals after subtraction of both the spot and planet model. Horizontal black lines show the phase of transit ingress (bottom) and egress (top). Vertical black lines are -$v$ sin $i$ (left) and  +$v$ sin $i$ (right).}
        \label{fig:trace}
\end{figure*}

A dark spot feature was seen during both spectroscopic transits on 2021-05-14 and 2021-06-18 (Figure~\ref{fig:trace}). We interpret this to be the same spot feature. That it reappears at the same location on the second transit is interpreted to be that the planet orbital period is a multiple of the rotation period, near 5:1 resonance. This is supported by the near-sinusoidal light curve variability seen during Sector 38 of the TESS observations. We therefore adopt a fixed stellar rotation period of $P_\mathrm{rot} = 1.39$\,d for both the Gaussian Process detrending and our Doppler tomographic analysis, and assume that there is one fixed long lived spot that does not vary in size or temperature between the two observations. Not taking into account the spot evolution explains the slight residuals left after removal of the main spot feature as seen on the bottom plot of Figure~\ref{fig:trace} (panels (a)-(b) and (a)-(b)-(c)). For reference, a Lomb-Scargle analysis of the joint Sector 11 and 38 light curves reveal a rotation period of $1.4\pm0.1$\,d, in agreement with the value we adopt in this analysis. 

\subsection{Global fit} 
\label{subsec:modeltotal}
We simultaneously fitted the transit light curves modeled with \textsc{batman} (Section~\ref{subsec:transitsphoto}) to the TESS data and the line profiles obtained through our Doppler tomographic model to the least squares deconvolution profiles (Section~\ref{subsec:transitsspectro}). \\
For a given parameter set, we computed the likelihood for both the photometric and spectroscopic datasets. To determine the best fitting parameter values and resulting posterior, we made use of a Markov chain Monte Carlo analysis using the affine-invariant ensemble sampler \textsc{emcee} \citep{2013PASP..125..306F}. The analysis included 240 simultaneous chains with 10000 iterations each. The resulting posteriors are presented in Table~\ref{tab:priors}. \\
We derived a planet radius of $10.178\pm0.44$ \REarth, agreeing with the previous value of  $10.07\pm0.47$ \REarth\,from  \citet{2020AJ....160...33R}. We note that the availability of four extra transits did not improve the uncertainty on $R_P$, limited by the poorer constraint on $R_{\star}$. \\

\begin{deluxetable*}{lll}
\tablenum{1}
\tablecaption{Global model parameters. \label{tab:priors}}
\tablewidth{0pt}
\tablehead{
\colhead{Parameters} & \colhead{Prior\tablenotemark{a}} & \colhead{Results}
}
\decimals
\startdata
\textbf{Transit} & & \\
$T_c$ (\TBJD) & $\mathcal{U}$ & 1604.023722 $_{-0.00023}^{+0.00024}$\\
$P_{\mathrm{orb}}$ (days) & $\mathcal{U}$ & 6.959471 $_{-0.0000030}^{+0.0000030}$\\
$R_\mathrm{p}$ (\Rstar) & $\mathcal{U}$ & 0.067471 $_{-0.00017}^{+0.00019}$\\
$a$ (\Rstar) & $\mathcal{U}$ & 11.685 $_{-0.220}^{+0.112}$\\
$i$ (\degree) & $\mathcal{U}$[86,90] & 89.23 $_{-0.47}^{+0.37}$\\
$e$ & 0 (fixed) & -\\
$\mu_1$ &  0.148 (fixed) \tablenotemark{b}  & -\\
$\mu_2$ &  0.23 (fixed)  \tablenotemark{b}& -\\
\hline
\textbf{Doppler tomography} & & \\
$v$ sin $i_\star$ (\kms)& $\mathcal{G}$[50,1] & 49.21 $_{-0.97}^{+0.95}$\\
$\lambda$ (\degree) & $\mathcal{U}$ & -5.8 $_{-5.7}^{+2.8}$\\
$v_{\mathrm{macro}}$ (\kms) & $\mathcal{U}$ & 0.59 $_{-0.41}^{+0.43}$\\
$\nu_1$ &  0.4139 (fixed) \tablenotemark{c}  & -\\
$\nu_2$ &  0.2494 (fixed)  \tablenotemark{c}& -\\
$\#_{\mathrm{spots}}$ &  1 (fixed) & -\\
$\theta_{\mathrm{spot1}}$ (\degree) & $\mathcal{U}$ & 212.95 $_{-0.37}^{+0.36}$\\
$\phi_{\mathrm{spot1}}$ (\degree) & $\mathcal{U}$[0,90] & 57.8 $_{-2.2}^{+2.7}$\\
$R_{\mathrm{spot1}}$ (\degree) & $\mathcal{U}$ & 4.76 $_{-0.71}^{+0.24}$\\
$T_{\mathrm{spot1}}$ (K) & $\mathcal{U}$[0.6$\times$Teff,0.8$\times$Teff] & 3890 $_{-1550}^{+288}$\\
$P_\mathrm{rot}$\,(days) & 1.39 (fixed) & -\\
\hline
\textbf{Lightcurve Gaussian Process} & & \\
$\log S_0$ & $\mathcal{U}$ & $-1.6_{-3.7}^{+0.4}$ \\
$\log Q$ & $\mathcal{U}$ & $3.37_{-0.77}^{+0.67}$\\
$\log \omega_0$ & -0.329 (fixed) & -\\
\hline
\textbf{Derived parameters} & & \\
$R_P$ (\REarth) & - & $10.178\pm0.440$\\
$R_P$ (\RJup) & - & $0.928\pm0.040$ \\
3D obliquity (\degree) & - & $20.2^{+10.3}_{-8.7}$ \\
$i_\star$ (\degree) & - & $>$ 85 (3$\sigma$) \\
\hline
\enddata
\tablenotetext{a}{$\mathcal{U}$ unconstrained uniform priors; $\mathcal{U}$ [$a,b$] uniform constrained priors with boundaries $a$ and $b$; $\mathcal{G}$[$\mu$,$\sigma$] Gaussian priors}
\tablenotetext{b}{Adopted at the TESS band from \citet{2017AA...600A..30C}}
\tablenotetext{c}{Adopted at the $V$ band from \citet{2012AA...546A..14C}}

\end{deluxetable*}

\section{Conclusions} \label{sec:discussion}

\begin{figure}
    \centering
    \includegraphics[width=1\linewidth]{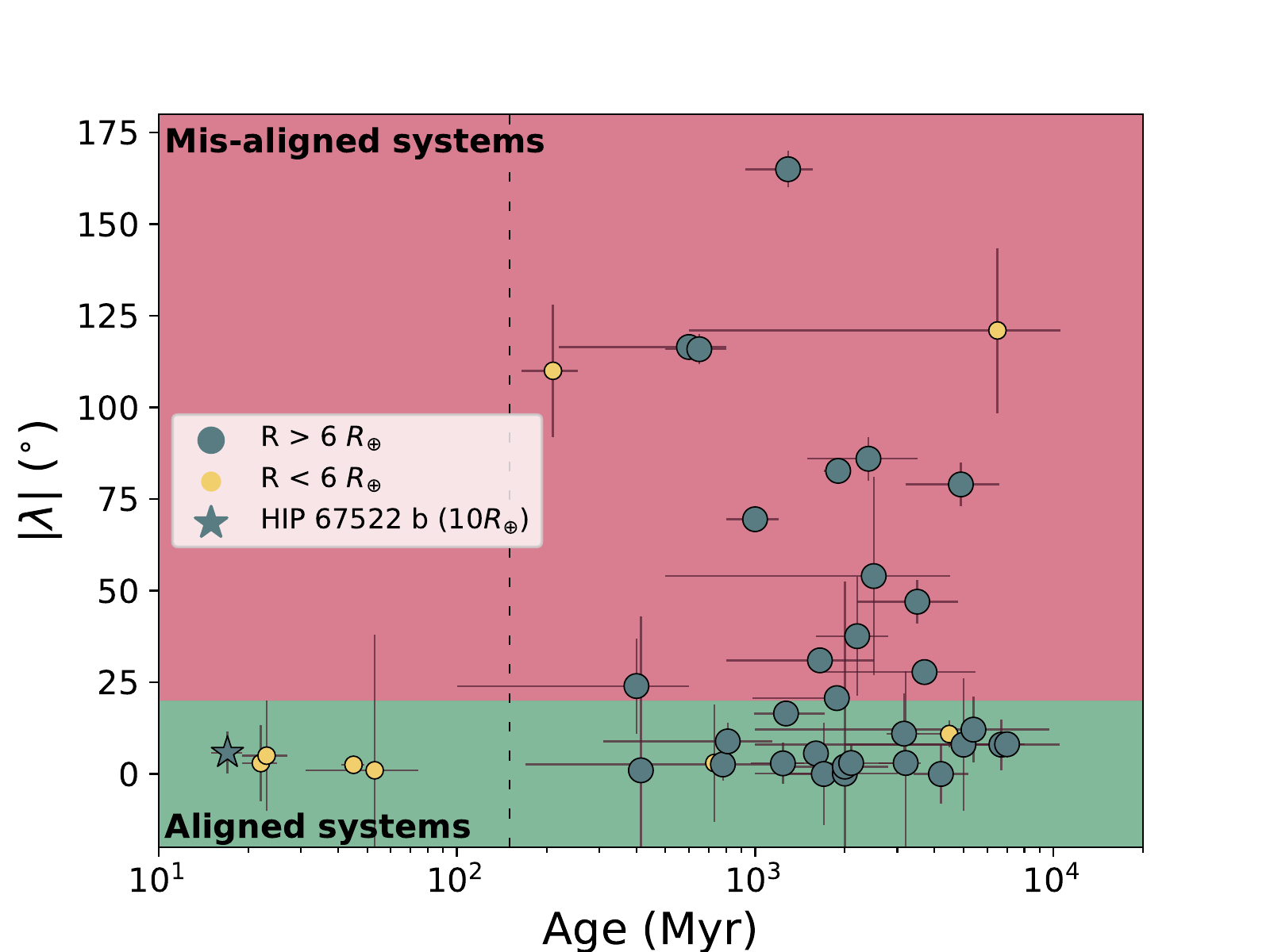}
    \caption{Systems younger then 10 Gyr that have both their age (NASA exoplanet archive 2021 July) and obliquity constrained. Yellow smaller circles represent Neptune sized planets ($R < 6$ \REarth), larger blue circles show Jupiter sized planets ($R > 6$ \REarth) and the blue star symbolizes HIP 67522 b. The very young systems ($< 100$ Myr) are, in ascending order of age, HIP 67522, AU Mic, V 1298 Tau, DS Tuc A and TOI 942. The red and green areas show aligned ($|\lambda| < 20^{\circ}$) and misaligned ($|\lambda| > 20^{\circ}$) systems respectively.}
    \label{fig:age-lambda-plot}
\end{figure}

We measured the projected obliquity angle of HIP 67522 b to be $|\lambda| = 5.8 ^{+2.8}_{-5.7}$\degree. With a stellar inclination estimated following \citet{2020AJ....159...81M}, we derived the 3D obliquity to be $\psi= 20.2^{+10.3}_{-8.7}$\degree. At an age of 17 Myr, HIP 67522 b is the youngest planet to receive such characterization. We demonstrate that a precise measurement of the sky-projected obliquity is possible for such young stars, despite the activity-dominated spectroscopic transit observations. Our single-spot model allows us to unambiguously disentangle the planetary signature from the stellar activity.

Figure~\ref{fig:age-lambda-plot} places HIP 67522 b into context of other planetary systems which have known obliquities and constrained ages\footnote{NASA Exoplanet Archive July 2021}. This particular system joins AU Mic b \citep{2020A&A...643A..25P,2020arXiv200613675A,2020A&A...641L...1M,2020ApJ...899L..13H}, V1298 Tau c \citep{2019ApJ...885L..12D,2021arXiv210701213F}, DS Tuc Ab \citep{2019ApJ...880L..17N,2020ApJ...892L..21Z,2020AJ....159..112M} and TOI 942 b \citep{2021arXiv210614968W} in the group of systems younger than 100 Myr old that have their obliquity measured. Remarkably, all these systems have been found to be on well-aligned orbits, encouraging the pursuit of very young star obliquity measurements to confirm this trend. 

With a radius of 10\,\REarth, HIP 67522 b is the only hot, Jupiter-sized planet in this very young group of systems. Its future evolution is still unclear however, and the lack of constraint on its mass prevents us from definitively classifying it as a proto-hot Jupiter. Super-Earths and Neptune-sized planets commonly found around Sun-like stars can have a radii of $\sim 10$\,\REarth\,at the age of HIP 67522. The planet is undergoing Kelvin-Helmoholz contraction and photoevaporation, and its eventual radius depends strongly on its core-envelope makeup \citep[e.g.][]{2013ApJ...776....2L}. 

If HIP 67522 b is indeed a proto-hot Jupiter, it is a prime example of one that did not migrate via high eccentricity pathways. The circularization of HIP 67522 b's orbit is on the Gyr timescale (estimated from \citealt{1966Icar....5..375G}), playing against a high eccentricity type migration. Classic planet systems such as HD 80606 b \citep{2001A&A...375L..27N,2003ApJ...589..605W,2009A&A...502..695P,2010A&A...516A..95H} and HD 17156 b \citep{2007A&A...476L..13B,2008PASJ...60L...1N,2008ApJ...683L..59C} exhibit highly eccentric and oblique orbits. For most other mature hot Jupiters though, it is more difficult to decipher their original migration pathways due to tidal synchronization that occurs at the hundreds of millions of years to gigayear timescales \citep[e.g.][]{2012MNRAS.423..486L,2014ApJ...786..102V}, erasing evidence of their primordial histories. HIP 67522 b may be the first such example for which the most likely explanation is that the primordial orbit of a close-in Jovian planet is well aligned. Recently, \citet{2021ApJ...916L...1A} found that hot Jupiters are preferentially found in well aligned or polar orbits. There are a range of mechanisms that can result in such bimodality in the obliquity distribution, and understanding the age-obliquity distribution can help distinguish between these mechanisms. For example, magnetic warping of the protoplanet disk can result in hot-Jupiters that formed in-situ being found in oblique orbits \citep[e.g][]{2011MNRAS.412.2790L}. With more observations, a prevalence of well aligned hot Jupiters around young stars may help limit the real-life effectiveness of such pathways. 

If HIP 67522 b does become a Neptune-sized planet, it adds to the well aligned pool of very young systems, contrasting with the often misaligned single-planet Neptune-like systems in close-orbit around more mature aged stars. Even with a mass of a few 10 \MEarth, the circularization timescale is likely on the few 100 Myr timescale, one order of magnitude older larger than the age of the system, disfavoring high eccentricity migration. HIP 67522 b would contribute to the growing interest to understand the formation of close-in Neptunes (e.g  \citealt{2021JGRE..12606639B}) and to make sense of the alignment distribution of these commonly found exoplanets.

Although not seen in sector 38, HIP 67522 b has a possible nearby exterior transiting companion with an orbital period of $>23$\,days \citep{2020AJ....160...33R}. The existence of HIP 67522 c is tentative, but if confirmed, HIP 67522 b would be more consistent future Neptune-sized planet as hot Jupiters are rarely found with outer companions. Dynamical interactions within closely packed planetary systems can excite mutual inclinations \citep{2013ApJ...775...53H}, and often result in the destruction of the interior planetary architecture if the outer companion is massive \citep{2017AJ....153..210H}. Systems like HIP 67522 and V1298 Tau \citep{2019ApJ...885L..12D,2021arXiv210701213F} are candidates to test planet-planet interactions before planetary systems have settled into their final stable forms.

\begin{acknowledgments}
This research has been supported by an Australian Government Research Training Program Scholarship.
GZ thanks the support of the ARC DECRA program DE210101893.
GZ thanks the support of the TESS Guest Investigator Program G03007.
This research has used data from the CTIO/SMARTS 1.5m telescope, which is
operated as part of the SMARTS Consortium by \href{http://secure-web.cisco.com/1TL5nionOJJUGi7T0X_YvX7RLRwbVQl20QG7s4LKeK1vpFY8M3UHYMuONVvV2D2hxli_pMi4YkHdTYel4ogZ3sJWN4axM8-5IsyCIPeIj7BfVIBOvp9a8iRKv2IM-wTBpjGA3xxZcH5lT4FNKBIoEstyJEEyUYzEKbDQyL4T1LQSiukl5eTarVlkS9YJbHf_HrjiuXV1gM1uXr7gdIdCbZg4CfJa_N8Qw38oz0KhpJ74RZ0rIcyg3XKCc6-HCDjlBrMtX3cpMKa1Kcya1SxY0UxXY0WkwM0zGeXYUYfbkp1Ce6jIBY8Evcz-YcyODRE4QWMlPqSDV66bKv5F1R3-RrkcH91Y7INyFOP6qJfGJKLRFJT-KNphpqmNc4Pf7zLVOIBjCEKsANmt1XTtzQN5AIPwKf-F1qd4b6KCZrqjHZIA/http\%3A\%2F\%2Fwww.recons.org}{RECONS}
members Todd Henry, Hodari James, Wei-Chun Jao, and Leonardo Paredes.  At
the telescope, observations were carried out by Roberto Aviles and Rodrigo
Hinojosa. In the scope of this research, we used the University of Southern Queensland's (USQ) \href{https://www.usq.edu.au/hpc}{Fawkes HPC} which is co-sponsored by the Queensland Cyber Infrastructure Foundation (QCIF).
This research has made use of the NASA Exoplanet
Archive, which is operated by the California Institute of Technology,
under contract with the National Aeronautics and Space Administration
under the Exoplanet Exploration Program. 
Funding for the TESS mission is provided by NASA's Science Mission directorate. We acknowledge the use of public TESS Alert data from pipelines at the TESS Science Office and at the TESS Science Processing Operations Center. This research has made use of the Exoplanet Follow-up Observation Program website, which is operated by the California Institute of Technology, under contract with the National Aeronautics and Space Administration under the Exoplanet Exploration Program. This paper includes data collected by the TESS mission, which are publicly available from the Mikulski Archive for Space Telescopes (MAST).
Resources supporting this work were provided by the NASA High-End Computing (HEC) Program through the NASA Advanced Supercomputing (NAS) Division at Ames Research Center for the production of the SPOC data products.
\end{acknowledgments}

\vspace{5mm}
\facilities{CTIO:1.5m, TESS}

\software{astropy \citep{2013A&A...558A..33A,2018AJ....156..123A}, batman \citep{2015PASP..127.1161K}, celerite \citep{2017AJ....154..220F}, emcee \citep{2013PASP..125..306F}}

\bibliography{main}{}
\bibliographystyle{aasjournal}

\end{document}